\documentclass[reprint, amsmath,amssymb, showkeys, aps]{revtex4-1}

\usepackage{graphicx}
\usepackage{dcolumn}
\usepackage{bm}

\newcommand{\Comment}[1]{{}}
\usepackage{amsfonts,amsthm,amsmath,amssymb,slashed}
\usepackage[textwidth = 430 pt, textheight = 630 pt]{geometry}

\Comment{\usepackage{color}
\definecolor{MyDarkBlue}{rgb}{0.15,0.15,0.45}
\usepackage[linktocpage=true]{hyperref}
\hypersetup{
colorlinks=true,
citecolor=MyDarkBlue,\bibliography{myref}
linkcolor=MyDarkBlue,
urlcolor=MyDarkBlue,
pdfauthor={Thiago Araujo},
pdftitle={DRAFT},
pdfsubject={hep-th}
}

\usepackage[numbers,sort&compress]{natbib}
\usepackage{hypernat}}
\usepackage{graphicx}

\newcommand\ignore[1]{}
\def\one{{\,\hbox{1\kern-.8mm l}}}

\def\a{\alpha}

\def\G{\Gamma}

\def\s{\sigma}

\def\d{\partial}

\newcommand{\Cset}{{\,\,{{{^{_{\pmb{\mid}}}}\kern-.45em{\mathrm C}}}}}

\newcommand{\be}{\begin{equation}}
\newcommand{\bea}{\begin{eqnarray}}

\newcommand{\ee}{\end{equation}}
\newcommand{\eea}{\end{eqnarray}}

\usepackage{color}
\newcommand{\bse}{\begin{subequations}}
\newcommand{\ese}{\end{subequations}}

\begin{document}

\title{Revisiting Wilson loops for nonrelativistic backgrounds}

\author{Thiago R. Araujo}
 \email{taraujo@ift.unesp.br}
 
\affiliation{
Instituto de F\'{i}sica Te\'{o}rica, UNESP-Universidade Estadual Paulista\\
R. Dr. Bento T. Ferraz 271, Bl. II, Sao Paulo 01140-070, SP, Brazil
}

\date{\today}

\begin{abstract}
We consider several configurations that describe Wilson loops in nonrelativistic field theories, and for some of them we find systems of coupled nonlinear differential equations. Also, we find a nontrivial drag force at zero temperature, which suggests that the parameter controlling the deviation of the nonrelativistic space from the relativistic space may be related to the chemical potential of these systems. Moreover, we reconsider some known configurations in the literature and we perform further analysis.
\end{abstract}

\keywords{Gauge/gravity duality; Wilson loops; holography; nonrelativistic; Schr\"odinger; Lifshitz}

\maketitle

\section{Introduction}

The general idea of the gauge/gravity conjecture is that a field theory in a space of $d$-dimensions can be equivalent to a gravity theory in $d+1$ dimensions, when the symmetries of the field theory are realized as isometries of the gravity side \cite{Maldacena:1997re, Witten:1998qj}. This innocent, but powerful, idea of the holographic principle has driven the vanguard of physics for almost twenty years.

Roughly speaking, the conjecture states that there is a correspondence between physical quantities in the gravitational side and physical quantities in the field theory side. One important quantity in the gauge field theory side is the \emph{Wilson loops}, which is a gauge invariant observable constructed from the connection, and is associated to the parallel transport of a particle moving through the gauge field \cite{Smilga2001, Makeenko:2009dw}.

In the holographic context, the prescription for the calculation of Wilson loops in the gravity side was given by \cite{Maldacena:1998im} and was applied in the $AdS_5\times S^5$ solution of type IIB supergravity which is dual to a conformal field theory. This prescription has been extended to backgrounds which are not anti-de Sitter, and by consequence, to backgrounds that are not dual to conformal field theories - although, they preserve Lorentz symmetry - see \cite{Drukker:1999zq, Sonnenschein:1999if, Nunez:2009da} for excellent reviews and further analysis.

Furthermore, another important aspect of the gauge/gravity duality is that it relates the large N and strong coupling regime of the field theory with weakly coupled gravitational theory. As a result, the savage strongly coupled regime of the field theory can be mapped to a docile weakly coupled regime in the gravitational side and vice-versa \cite{Maldacena:1998im, Aharony:1999ti, Polchinski:2010hw}. 

Since we have a plethora of strongly coupled systems in condensed matter physics, it is perfectly reasonable to look for a gravitational dual to theories which describe these condensed matter systems. Then, it is clear now that we have a new paradigm in the gauge/gravity conjecture, namely, the field theories in condensed matter fields are nonrelativistic, so we need to consider that their dual backgrounds have nonrelativistic isometries \cite{Gomis:2000bd, Nishida:2007pj, Son:2008ye, Balasubramanian:2008dm, Herzog:2008wg, Maldacena:2008wh, Adams:2008wt, Ko:2015rha}, see \cite{Hartnoll:2009sz} for a review.

The holography for nonrelativistic systems is at an incipient stage, and there are several unknown aspects that we need to understand, for example, in theories that exhibit a $d$-dimensional Schr\"odinger symmetry, their algebra cannot be organized as an isometry of a $(d+1)$-dimensional space as usual, but in a $(d+2)$-dimensional space, and the role of this extra dimension is still unclear. 

Moreover, just recently the gravity duals of some of these relativistic systems have been embedded into string theory, see for instance \cite{Balasubramanian:2008dm, Donos:2009xc, Donos:2010tu, Balasubramanian:2010uk, Gregory:2010gx}, and it is evident that the fundamental nature of the field theories is still a mystery. However, one may hope to be able to identify and elucidate aspects of the nonrelativistic dual field theories, just applying the holographic principle,  in the gravity side, in the calculation of familiar physical quantities.

Wilson loops seem to be a good starting point, since it is related to a probe string moving just on the external space. This means that we can ignore, for a moment, the internal space (as well as additional fields, such as the dilaton and p-forms) which composes the supergravity solution. 

In this sense, it is an observable which demands a small amount of information about the background where the string is moving in, but it gives us important information about the nature of the field theory; for instance, if the theory confines, if the theory has conformal symmetry and so on \cite{Sonnenschein:1999if, Nunez:2009da}. Also, we can compute drag forces and the energy loss of charged particles moving in these backgrounds \cite{Gubser:2006bz, Chernicoff:2006hi, Herzog:2006se, Herzog:2006gh}.

In this work we consider the calculation of holographic Wilson loops on backgrounds with nonrelativistic symmetries. As a word of caution, one needs to remember that as pointed in \cite{Hartnoll:2009sz}, we should be careful in using spaces with nonrelativistic symmetries, mainly if we want to consider regions near the end of the space. In the IR we can find null singularities on these spaces, which can be removed as
soon as we consider finite temperature effects \cite{Hartnoll:2009sz, Adams:2008wt, Horowitz:1997uc}.

We start with a short review of the usual prescription of Wilson loops defined in \cite{Maldacena:1998im}, reviewed in \cite{Sonnenschein:1999if}. Also, we highlight the relevant details for the calculation of the drag forces considered in \cite{Gubser:2006bz, Herzog:2006gh}.

In the main part of this text, we examine some string configurations on backgrounds with Schr\"odinger and Lifshitz symmetries and we see that these systems are tricky. We exclude some configurations and we also find systems that can hardly be solved analytically.

Even though the nonrelativistic systems considered here are at zero temperature, we found a nonzero drag force for them, as in \cite{Akhavan:2008ep}. Finally, reconsidering the systems of \cite{Kluson:2009vy, Siahaan:2011sw}, we perform further analysis and present some speculative ideas on the nature of the nonrelativistic field theory dual to the background.

\section{Short review}

In this section we present a short review of the ideas examined in \cite{Maldacena:1998im, Kinar:1998vq} and reviewed in \cite{Sonnenschein:1999if, Nunez:2009da} for the calculation of the quark-antiquark distance and potential. In addition, we want to consider some fundamental ideas related to the drag force on a classical string configuration as in \cite{Gubser:2006bz, Chernicoff:2006hi, Herzog:2006gh}.

\subsection{Quark-antiquark system}

We start with a background of the generic form
\be 
ds^2=-g_{tt}dt^2+g_{xx} d\vec{x}^2+g_{rr}dr^2+ds^2_{\cal M}
\ee
where $g_{tt}, g_{xx}$ and $g_{rr}$ are functions of the radial coordinate $r$, and the term $ds^2_{\cal M}$ is a metric of an internal manifold. We can neglect the internal space $ds^2_{\cal M}$ because we consider a probe string that is not excited along those directions.

We take an ansatz for the string as 
\be 
t=\tau\; , \quad x=x(\s)\; , \quad r=r(\s)\; ,
\ee
and when we calculate the Nambu-Goto action and its equations of motion, we find that this configuration implies
\be 
\frac{dr}{d\s}=\pm \frac{dx}{d\s}\frac{f(r)}{C_0 g(r)}\sqrt{f(r)^2-C_0^2}\; ,
\ee
where $f(r)^2=g_{tt}g_{xx}$, $g(r)^2=g_{tt}g_{rr}$ and $C_0$ is an integration constant. The shape of the solution in this background can be pictured as a string whose ends are fixed at $x = 0$ and $x = \ell_{q\bar{q}}$ at the boundary of space, $r\to 0$. In addition, it can extend in the bulk, so that the radial coordinate of the string assumes its maximum value at $r_0$, that occurs at $x = \ell_{q\bar{q}}/2$. Furthermore, one can show that the integration constant is equal to $C_0=f(r_0)$, see \cite{Nunez:2009da}.

Considering the string solution above, we can compute gauge invariant quantities such as the separation and the energy between the endpoints of the string, which can be interpreted as the separation between a quark and an antiquark living on the brane, see \cite{Sonnenschein:1999if, Nunez:2009da} for further details. These results are given by
\begin{widetext}
\begin{align}
\ell_{q\bar{q}}(r_0)&= 2f(r_0)\int_0^{r_0} dr \frac{g(r)}{f(r)}\frac{1}{\sqrt{f(r)^2-f(r_0)^2}}\; ,\\
E_{q\bar{q}}(r_0)&=f(r_0)\ell_{q\bar{q}}(r_0)-2 \int_0^{r_0} dr g(r)
 +2 \int_0^{r_0} dr \frac{g(r)}{f(r)}\sqrt{f(r)^2-f(r_0)^2}\; .
\end{align}
\end{widetext}

\subsection{Drag force}\label{drag-force}

In \cite{Gubser:2006bz}, the author considered a probe string moving through the AdS${}_5$-Schwarzschild background, whose radius of the horizon is related to the temperature of the dual gauge theory. In summary, Gubser considered the metric of the near-extremal D$3$-brane
\be 
ds^2_{10}=\frac{(-h dt^2+d\vec{x}^2)}{\sqrt{H}}+\sqrt{H}(dr^2/h+d\Omega_5^2)
\ee
where 
\be 
H=1+\frac{L^4}{r^4}\; , \quad h=1-\frac{r_H^4}{r^4}.
\ee
The near horizon limit is simply 
\be 
ds^2=\frac{r^2}{L^2}(-h dt^2+d\vec{x}^2)+\frac{L^2}{r^2}\frac{dr^2}{h}\; ,
\ee
where we drop the five dimensional part of the metric, since it plays no role in the present case.

Besides, he considered the following configuration
\be 
t=\tau\; , \quad x(\tau, \s)=v_x\tau + \eta(\s)\; , \quad r=\s\; ,
\ee
with action
\be 
S=\frac{1}{2\pi \alpha'}\int d\tau d\s {\cal L}\; ,
\ee
and density
\be 
{\cal L}=\sqrt{1-\frac{v_x^2}{h}+\frac{h}{H}\eta'^2}\; .
\ee
From the equation of motion we find that the momentum $\Pi_\eta=\frac{\d {\cal L}}{\d \eta'}$ is a constant equals to
\be 
\Pi_\eta=\frac{v_x}{\sqrt{1-v_x^2}}\frac{r_H^2}{L^2}\; .
\ee
Using this last expression, the authors of \cite{Herzog:2006gh} showed that the drag force, opposite to the motion of the string, is given by 
\be 
F_{\eta}=-\frac{1}{2\pi \alpha'}\Pi_{\eta}, 
\ee
and using the relation $\pi L^2 T=r_H$, we see that the drag force depends on the temperature of the system.

In this section we have defined the calculation of the drag force using the holographic principle, but we can reconsider this same calculation for backgrounds without horizon. This is what we intend to do below.

\section{Schr\"odinger backgrounds}

We begin by reconsidering some of the calculations that have been performed in \cite{Akhavan:2008ep, Kluson:2009vy,  Siahaan:2011sw, Araujo:2015dba} concerning the calculation of Wilson loops on backgrounds with Schr\"odinger symmetries. Moreover, we perform further analysis in these solutions, and we study additional string configurations.

The probe string moves on a manifold of the form 
\be 
ds^2=\frac{R^2}{r^2}\left( \frac{-dt^2}{r^{2(z-1)}}+2 d\xi dt + (dx^i)^2 +dr^2\right) \label{metric-schr}
\ee
where $\xi$ is a compact timelike coordinate, and the natural number $z$ is the dynamical exponent. It can be shown , see \cite{Hartnoll:2009sz} and references therein, that for $i=1, \cdots, D-1$, the space (\ref{metric-schr}) is the geometric realization of the Schr\"odinger algebra in $D$ dimensions.

\subsubsection{\bf Constant compact direction} \label{schr-const}

First, we consider the following configuration for the probe string \cite{Siahaan:2011sw, Kluson:2009vy, Akhavan:2008ep}
\be 
t= \tau\; , r=r(\s)\; , x=x(\s)\; , \xi=constant\; .
\ee
The Nambu-Goto action for this configuration is
\be 
{\cal S}=\frac{T}{2\pi \a'}\int d\s \sqrt{\frac{R^4}{r^{2(z+1)}}\left((x')^2+(r')^2 \right)}\; ,
\ee
and if we define $f(\s)=R^2/r^{(z+1)}$, the equations of motion for $x$ and $r$ are
\begin{align} 
\phantom{\d_\s \left(\right)}\frac{f^2 x'}{\sqrt{f^2(x'^2+r'^2)}}&=C_0\label{feq_mot}\\
\d_\s\left(\frac{f^2 r'}{\sqrt{f^2(x'^2+r'^2)}}\right)&=\frac{ (x'^2+r'^2)}{\sqrt{f^2(x'^2+r'^2)}}f\frac{df}{dr} \label{seq_mot}.
\end{align}
Equation (\ref{feq_mot}) implies that
\be 
\frac{dr}{d\s}=\pm \frac{dx}{d\s}\; \frac{\sqrt{f^2-C_0^2}}{C_0}=\pm \frac{dx}{d\s}\;V_{eff}(r)\; ,
\ee
and the equation (\ref{seq_mot}) is solved when this last equation is satisfied. Also, for a $\cap$-shaped string, the turning point is defined as the point $r_0$ where $\left. \frac{d r}{d x}\right|_{r_0}=0$. Using this condition we determine the constant $C_0=f(r_0)$.

Since we consider a string moving in the bulk with its endpoints lying on the boundary $r\to 0$, the Dirichlet boundary condition must be satisfied, that is $\lim_{r\to 0} \frac{dx }{dr}\to 0$. We can see that this condition is readily satisfied, since $\lim_{r\to 0}V_{eff}\to \infty$.

The quark-antiquark distance is  given by
\be
\ell_{q\bar{q}}(r_0, z)=2 r_0 \sqrt{\pi} \frac{\G\left(\frac{z+2}{2(z+1)}\right)}{\G\left(\frac{1}{2(z+1)}\right)},
\ee
and in \cite{Kluson:2009vy, Siahaan:2011sw} it was found that the quark-antiquark potential \footnote{Using the definition of the Gamma function to extend the domain of the Beta function.} is
\begin{widetext}
\be
\begin{split}
V_{q\bar{q}}(r_0, z)&=-2\frac{ R^2 \sqrt{\pi}}{z r_0^z }
\frac{\G\left(\frac{z+2}{2(z+1)}\right)}{\G\left(\frac{1}{2(z+1)}\right)}\\
&=-\frac{(2\sqrt{\pi})^{1+z} R^2}{z }
\left(\frac{\G\left(\frac{z+2}{2(z+1)}\right)}{\G\left(\frac{1}{2(z+1)}\right)}\right)^{z+1}\frac{1}{\ell_{q\bar{q}}^z}\; .
\end{split}
\ee
\end{widetext}
From this equation we see that for the special case $z=1$ we have the behaviour $V_{q\bar{q}}\sim -\frac{1}{\ell_{q\bar{q}}}$ in the potential quark-antiquark, which is consistent with the conformal scaling.

In \cite{Siahaan:2011sw}, the author also showed that the convexity conditions \cite{Bachas:1985xs, Arias:2009me} of such a configuration are satisfied, that is
\be 
\frac{dV_{q\bar{q}}}{d\ell_{q\bar{q}}}>0\; , \quad \frac{d^2V_{q\bar{q}}}{d\ell_{q\bar{q}}^2}\leq 0,
\ee
where the first condition means that the quark-antiquark interaction is always attractive and the second equation means that the potential is a monotone nonincreasing function of $\ell$. Therefore, this configuration is physically admissible.

A second configuration with constant compact direction that we would like to explore is given by
\be 
t= \tau\; , r=\s\; , x=v_x \tau+\eta(\s)\; , \xi=const\; .
\ee
As we said in the previous section, the drag force has been studied in \cite{Gubser:1998bc,Herzog:2006se} in the context of a quark moving in a thermal plasma of ${\cal N}=4$ SYM, and we have seen that the horizon is related to the temperature of the field theory. Despite the fact that in the present case we do not have a horizon in our geometry, we may apply the very same ideas.

The action is
\be 
{\cal S}=\frac{T}{2\pi \a'}\int dr \sqrt{\frac{R^4}{r^2}\left[\frac{1}{r^{2z}}+\frac{\eta'^2}{r^{2z}}-\frac{v^2}{r^{2}} \right]}\; ,
\ee
and the equation of motion implies that $\Pi_\eta$, given by 
\be 
\Pi_\eta=\frac{R^4 \eta'}{r^{2z+2}\sqrt{\frac{R^4}{r^2}\left[\frac{1}{r^{2z}}+\frac{\eta'^2}{r^{2z}}-\frac{v^2}{r^{2}} \right]}}\; ,
\ee
is a constant. Therefore, we find 
\be 
\eta'=\Pi_\eta \sqrt{\frac{1-v^2 r^{2z-2}}{\frac{R^4}{r^{2z+2}}-\Pi_\eta^2}}\; .\label{eta-func}
\ee
Now observe that if we take $z=1$, the numerator in the square root is positive for all values $v<1$, and this is consistent with a relativistic theory. 

For the denominator we find that for some large $r_\ast$ the constant $\Pi_\eta^2$ could be greater than $R^4/r^4_\ast$, and in this case, the denominator would be negative. Since there is no upper bound for $r$, we see that the reality condition of the integral implies that $\Pi_\eta=0$, which implies that the drag force is zero, as we shall see below. 

This result is expected, since for $z=1$ we have the anti-de Sitter space, which is at zero temperature, and in the relativistic case, the drag force for a system at zero temperature vanishes. Also, we can see that the equation of motion for $r$ is trivially satisfied since $\left.\frac{\d {\cal L}}{\d r'}\right|_{r=\s}={\cal L}$.

For $z=2$, the values $\Pi_\eta=\pm R^2 v^3$ avoid an imaginary value in (\ref{eta-func}). Essentially these two examples were studied in \cite{Akhavan:2008ep}. 

In addition, for $z>1$ we have the general formula 
\be 
\Pi_\eta= \pm R^2 v^{(z+1)/(z-1)}\; .
\ee
Using that the drag force, formally defined as
\be 
F_{drag}=-\sqrt{-g}G_{xx}g^{\s\s}\eta'\; ,
\ee
and that $\Pi_\eta=\frac{\d \cal{L}}{\d \eta'}$, we can easily show that $F_{drag}=\Pi_\eta$. The drag force is defined to be contrary to the velocity of the string, hence
\be 
F_{drag}=- R^2 v^{(z+1)/(z-1)}\; .
\ee
For the special case $z=2$, the drag force is $F_{drag}=-R^2 v^{3}$, which is consistent with the results found in \cite{Akhavan:2008ep}. 

For a general $z$, the results above are equal to the case studied in \cite{Fadafan:2009an} for the Lifshitz spacetime at zero temperature. This happens because any configuration in the Schr\"odinger and Lifshitz spacetimes have the same Nambu-Goto action when $\xi=constant$.

In the section \ref{drag-force} we have seen that the radius of the horizon is related to the drag force of the system. On the other hand, the nonrelativistic spaces we consider do not have horizons, but have nontrivial drag forces. As the authors argued in \cite{Akhavan:2008ep}, these systems may have a hidden chemical potential that allows such a phenomenon. In fact, making the transformations $t \to \mu t$ and $\xi \to \mu^{-1}\xi$ in (\ref{metric-schr}), we can repeat our calculations and see that $F_{drag}\propto 1/\mu^2$, and in the dual field theory, the parameter $\mu$ can be interpreted as the chemical potential \cite{Hartnoll:2009sz, Balasubramanian:2010uw}.

In other words, the chemical potential is the conjugate variable to the particle number, and the compact coordinate $\xi$ is directly related to the particle number (see for instance \cite{Hartnoll:2009sz, Balasubramanian:2010uw}); then it is somewhat expected the presence of this 'hidden' chemical potential. On the other hand, the nature of the coordinate $\xi$ is still a mystery \cite{Son:2008ye, Balasubramanian:2008dm, Herzog:2008wg, Maldacena:2008wh}, and the mechanism (considering that it exists) which allows us to relate the spectrum of the masses (particle number) to the chemical potential is unknown.

\subsubsection{\bf Nonconstant compact direction}

We now consider that the string also moves on the compact direction $\xi$. We start with an example studied by \cite{Kluson:2009vy}, where the author concluded that the configuration is not physical. Here we point out some reasons that suggest a richer physical scenario. Furthermore, we study a new configuration in which the compact direction $\xi$ depends on the coordinate $\s$ that parametrizes the string. This configuration is described by a system of nonlinear differential equations and we could not find an explicit solution. 

The reader must remember that we do not have a correct interpretation of this coordinate \cite{Hartnoll:2009sz}, consequently, the physical meaning of the string with its endpoints moving along this direction is uncertain; and maybe it is not even physically admissible. Even so, let us insist on this direction and examine the ansatz
\be 
t= \tau\; , r=r(\s)\; , x=x(\s)\; , \xi=\xi(\tau)\; ,\label{unp-ansatz}
\ee
where the Nambu-Goto action reads
\be 
{\cal S}=\frac{1}{2\pi \a'}\int d\tau d\s \sqrt{g(\tau, \s)^2\left((x')^2+(r')^2 \right)}\; ,
\ee
for $g(r,\xi)^2=\frac{R^4}{r^2}\left(\frac{1}{r^{2z}}-\frac{2}{r^2}\d_\tau \xi\right)$. The equations of motion are
\begin{align}
\d_\tau g(r, \xi)&=0\; ,\\ 
\phantom{\d_\s \left(\right)}\frac{g^2 x'}{\sqrt{g^2(x'^2+r'^2)}}&=C_1\; ,\\
\d_\s\left(\frac{g^2 r'}{\sqrt{g^2(x'^2+r'^2)}}\right)&=\frac{ (x'^2+r'^2)}{\sqrt{g^2(x'^2+r'^2)}}g\frac{dg}{dr} .
\end{align}
We can see that $\xi(\tau)=v_\xi \tau$, and the third equation is solved by imposing the second one. Therefore, the quark-antiquark distance now reads
\be 
\ell_{q\bar{q}}^\xi=\frac{2 g(r_0)}{R^2} \int_0^{r_0} \frac{dr\; r^{z+1}}{\sqrt{1-2v_\xi r^{2z-2}-\frac{g^2(r_0)}{R^4}r^{2z+2}}},
\ee
and the potential is
\begin{align}
V_{q\bar{q}}&=2\int_0^{r_0} \frac{ dr \ g(r)}{\sqrt{g^2(r)-g^2(r_0)}}-2\int_0^{\hat{r}_0} dr g(r), \label{poten}
\end{align}
where $\hat{r}_0$ is the end of the space. The last term is necessary to remove the infinity part of the potential \cite{Maldacena:1998im, Drukker:1999zq, Nunez:2009da, Sonnenschein:1999if}. This term is the mass of a W-boson which corresponds to strings stretching from zero to the end of the space $\hat{r}_0$. Additionally, the IR limit is defined such that the maximum value $r_0$ approaches the end of the space, that is $r_0 \to \hat{r}_0$ \cite{Nunez:2009da}. 

In \cite{Kluson:2009vy}, the author argued that since this integral is imaginary for values of $r$ such that $r^{2(z-1)}>r_\ast^{2(z-1)}=1/2v_\xi$, the configuration (\ref{unp-ansatz}) with $\xi=v_\xi\tau$ is unphysical. Even though his arguments seem accurate, we point out some reasons which suggest that, perhaps, it is too early to rule out this configuration, inasmuch as we must be careful in using nonrelativistic spaces in our calculations.

First we need to remember that, except when $z=1$, which is the \textit{AdS} space, we can have several undesirable features on the background such as curvature singularities at the end of the spacetime $\hat{r}_0\to \infty$, see \cite{Hartnoll:2009sz, Horowitz:1997uc, Blau:2009gd}. On the other hand, it is important to notice that not all curvature singularities affect physical quantities \cite{Gubser:2000nd, Charmousis:2010zz, Lippert:2014jma}, therefore, these spaces are not severely ill-defined. 

In fact, considering spaces with Lifshitz symmetry the authors of \cite{Andrade:2014bsa} considered a configuration that can be interpreted as scattering amplitudes and studied observable consequences of the singularity in the IR structure of the dual field theory. Moreover, we are considering zero temperature systems and these singularities can be removed with finite temperature effects \cite{Adams:2008wt, Hartnoll:2009sz}.

Therefore, we expect to integrate the counterterm in (\ref{poten}) up to some point $\hat{r}_0< \infty$. In this case, the problem can be fixed if we consider that the ``cutoff'' is defined at some point $\hat{r}_0<  r_\ast$, where the integral is well defined. 

Evidently, after fixing the end of the space $\hat{r}_0$, we have a maximum (allowed) value for the velocity $v_\xi$. Then, we can take a velocity $v_\xi$ to be small enough, such that $\hat{r}_0^{2(z-1)}<1/2v_\xi$. Therefore, we notice that for the case $v_\xi\neq 0$, we have a reasonable configuration under certain conditions, and also that the velocity along the compact direction $\xi$ may have an upper bound.

Furthermore, under the time reversal transformation $t\to -t$ or the parity $\xi \to -\xi$, the space (\ref{metric-schr}) is not invariant and we have $g^2=\frac{R^4}{r^2}\left(\frac{1}{r^{2z}}+\frac{2 v_\xi}{r^2}\right)$. In this case, the integral for the W-boson mass is always real. This is a hint that field theory dual to supergravity solutions with Galilean symmetries may ``perceive'' the time direction. This is a point that deserves further investigation.

Alternatively, since the coordinate $\s$ which parametrizes the string length is compact, we could consider a configuration with $\xi=\xi(\s)$, for $\s\in [0, L]$. Then, take the ansatz
\be 
t= \tau\; , r=r(\s)\; , x=v_x\tau + \eta(\s)\; , \xi=\xi(\s),
\ee
with Nambu-Goto action
\be
{\cal S}=\frac{T}{2\pi \a'}\int d\s {\cal L}\; ,
\ee
where
\be
\frac{{\cal L}^2}{R^4}=\left(\frac{1}{r^{2z}}-\frac{v_x^2}{r^2} \right)\frac{\left(\eta'^2+r'^2 \right)}{r^2}+\frac{\left(\xi' \right)^2}{r^4}\; ,
\ee
and equations of motion
\begin{equation}
\d_\s\left( \frac{R^4}{r^4{\cal L}}\xi' \right)=0\; ,
\end{equation}
\begin{equation}
\d_\s   \left[ \frac{R^4}{r^2{\cal L}} \left( \frac{1}{r^{2z}}-\frac{v_x^2}{r^2} \right)\eta' \right]=0\; ,
\end{equation}
\begin{equation}
\frac{\d {\cal L}}{\d r}=\d_\s\left[\frac{R^4}{r^2 {\cal L}}\left(\frac{1}{r^{2z}}-\frac{v_x^2}{r^2} \right)r'  \right]\; .
\end{equation}
The first two equations above give
\begin{widetext}
\begin{align}
\frac{1}{r^2}\left( \frac{R^4}{r^4}-C_1^2 \right)\xi'^2&= C_1^2\left(\frac{1 }{r^{2z}}-\frac{v_x^2}{r^2}\right)(\eta '^2+r'^2)\; ,\label{eq.1}\\
\frac{R^4}{r^4}\left(\frac{1}{r^{2z}}-\frac{v_x^2}{r^2} \right)^2\eta'^2&=C_2^2\left[\left(\frac{1}{r^{2z}}-\frac{v_x^2}{r^2} \right)\frac{(\eta'^2+r'^2)}{r^2}+\frac{\xi'^2}{r^4} \right]\; ,\label{eq.2}
\end{align}
\end{widetext}
respectively.

We can simplify this system considering the particular case $v_x=0$. The action is
\be
{\cal S}=\frac{T}{2\pi \a'}\int d\s \sqrt{h^2\left[\frac{(\eta')^2+(r')^2}{r^{2z}}+\frac{(\xi')^2}{r^2} \right]}\; ,
\ee
with $h(\s)^2=R^4/r^2$. Using the notation $2\pi \a' {\cal S}=T\int d\s {\cal L}$, we see that the equations of motion are
\begin{align} 
\d_\s\left(\frac{h^2}{\cal L}\frac{\xi'}{r^{2}} \right)&=0\; ,\\
\d_\s\left(\frac{h^2}{\cal L}\frac{\eta'}{r^{2z}} \right)&=0\; ,
\end{align}
and 
\be
\d_\s\left(\frac{h^2}{\cal L}\frac{r'}{r^{2z}} \right)=\frac{\d {\cal L}}{\d r}\; \label{m-eq} .
\ee
The first of these equations implies that
\be 
\frac{1}{r^2}\left( \frac{h^2}{r^2}-C_1^2 \right)(\xi')^2=\frac{C_1^2 }{r^{2z}}\left(\eta '^2+r'^2\right)\; ,\label{eq.3}
\ee
while the second equation gives
\be 
\frac{1}{r^{2z}}\left( \frac{h^2}{r^{2z}}-C_2^2 \right)(\eta ')^2=C_2^2 \left(\frac{r'^2}{r^{2z}}+\frac{\xi'^2}{r^{2}}\right)\; .\label{eq.4}
\ee
In order to find a restricted class of solutions, we consider initially $\xi=constant$, which is the first case considered in this section. Alternatively, we can set $\eta=constant$ and $r=\s$. In the latter, the equation of motion for $\eta$ is trivially satisfied, which means that $C_2=0$, whereas the equation of motion for $\xi$ gives
\be 
\frac{d r}{d \xi}=\pm \frac{r^{z-1}}{C_1}\sqrt{\frac{h^2}{r^2}-C_1^2}\; \label{eq1}.
\ee
The equation (\ref{m-eq}) now gives us the following differential equation
\be 
\frac{h(\xi')^2}{r^2}=-C_3 \sqrt{\frac{1}{r^{2z}}+\frac{ (\xi')^2}{r^2}}\; \label{eq2},
\ee
and if we insert (\ref{eq1}) into (\ref{eq2}) we see that this system is not consistent unless $C_1=C_3=0$, which implies that $\xi=constant$. This means that this restricted class of solutions is trivial, and in order to find solutions one may try to solve numerically the coupled equations (\ref{eq.1} -- \ref{eq.2}) or (\ref{eq.3} -- \ref{eq.4}) for $v_x=0$. We leave the numerical studies of the solutions in this paper for a future work.

In summary, one sees that the motion of string along the compact coordinate $\xi$ of a Schr\"odinger background is a tricky issue, and deserves further investigation, but in principle, there is no apparent reason to rule out these configurations.

\section{Lifshitz}

Now we would like to study the motion of a string in a space of the form
\be 
ds^2=\frac{R^2}{r^2}\left( -\frac{dt^2}{r^{2(z-1)}} + (dx^i)^2 \right)+\frac{R^2}{r^2}dr^2. \label{metric-lif}
\ee
Analogously to what we have done in the last section, we could try to consider the probe string with the following profile
\be 
t= \tau\; , r=r(\s)\; , x=x(\s)\; ,
\ee
and we get the same equations as in the first example of section (\ref{schr-const}), see \cite{Danielsson:2009gi}. Moreover, if we take the example
\be 
t= \tau\; , r=r(\s)\; , x=v_x \tau + \eta(\s) \; ,
\ee
we find the second example of the same section, since in that case we considered $\xi=constant$. 

Additionally, the solution presented in \cite{Donos:2010tu} is much more interesting. In this paper, the authors used the methodology of \cite{Donos:2009xc}, which allowed them to embed a Schr\"odinger invariant solutions with $z=2$ into string theory, to find a supergravity solution with Lifshitz symmetry.

The exterior part of the solution \cite{Donos:2010tu} is given by
\be 
ds^2=\frac{R^2}{r^2}\left(- 2 dt d\xi + (dx^i)^2 +r^2f(\xi)d\xi^2 +dr^2 \right) , \label{metric-lif2}
\ee
and in order to make explicitly the Lifshitz symmetry, we write
\be 
\frac{- 2 dt d\xi}{r^2} + d\xi^2\equiv-\frac{dt^2}{r^{4}} + \left(d\xi- \frac{dt}{r^2}\right)^2,
\ee
where we considered $f=1$.

Notice that a configuration with $\xi=constant$, $t=t(\tau)$, $r=r(\s)$ and $x=x(\s)$ is not allowed, since it would give a zero Nambu-Goto action. On the other hand, we may consider that
\be 
t= \tau\; , r=r(\s)\; , x=v_x\tau+\eta(\s)\; , \xi=const,
\ee
and we obtain a nontrivial Nambu-Goto action
\be
{\cal S}=\frac{T}{2\pi \a'}\int d\s \frac{R^2 v_x}{r^2} \sqrt{-(\eta')^2-(r')^2}\; .
\ee
From the reality of the action, we may notice that the functions $\eta$ and $r$ must be purely imaginary or one of them complex, in such a way that the combination $-(\eta')^2-(r')^2>0$. Such conditions for $\eta$ and $r$ are unacceptable because they are distances. Therefore, this configuration is unphysical.

In \citep{Araujo:2015dba}, we have studied one more configuration, namely
\be 
t= \tau\; , r=r(\s)\; , x=x(\s)\; , \xi=v_\xi\tau\; ,
\ee
and we saw that the the quark-antiquark distance is given by 
\be 
\ell_{q\bar{q}}(r_{max})=\frac{2\sqrt{2}}{\sqrt{f v_\xi}}\sqrt{-\frac{k^2+1}{k^2}}(\mathbf{K}(k)-\mathbf{E}(k))\; ,
\ee
where $\mathbf{K}(k)$ and $\mathbf{E}(k)$ are the complete elliptic integrals of first and second kind respectively; and $k^2=(fv_\xi r^2_{max}-2)/2$, with $-k^2\in (0,1)$. The quark antiquark potential is
\begin{widetext}
\be
\begin{split} 
V_{q\bar{q}}(r_{max})=\frac{L^2 \sqrt{2 v_\xi}}{r_{max}} \left[ -2\left(\frac{k^2 \sqrt{f v_\xi}}{2\sqrt{2}}\sqrt{\frac{-k^2}{1+k^2}} \ell_{q\bar{q}} +\frac{a}{2}\mathbf{E}(k)\right)\right. &-2\pi n \sqrt{2a}\\ 
&\left.+ \sqrt{2a} \arcsin \left(\sqrt{\frac{a}{2}}\right) + 2\sqrt{1-\frac{a}{2}}\; \right] .
\end{split}
\ee
\end{widetext}
where $a\in (0,2)$.

As we said before, the coordinate $\xi$ is compact, so it is reasonable to take the functional relation $\xi=\xi(\s)$. Then we consider the ansatz
\be 
t= \tau\; , r=r(\s)\; , x=v_x \tau+\eta(\s)\; , \xi=\xi(\s)\; ,
\ee
such that
\be
{\cal S}=\frac{T}{2\pi \a'}\int d\s {\cal L} \; ,
\ee
where
\be 
{\cal L}=R^2\frac{\sqrt{(1-r^2 v_x^2 f)(\xi')^2-v_x^2(\eta'^2+r'^2)}}{r^2}.
\ee
The equations of motion for $\eta$ and $\xi$ give
\bse 
\be 
\eta'=\pm \frac{C_0 r^2}{v_x}\sqrt{\frac{(1-r^2 v_x^2 f)\xi'^2-v_x^2 r'^2}{R^4v_x^2+C_0^2r^4}},
\ee
\be 
\frac{v_x^2 (\xi')^2 R^4 }{2r^2 {\cal L}}\d_\xi f+\d_\s\left(\frac{R^4(1-r^2 v_x^2f)\xi'}{r^4{\cal L}} \right)=0,
\ee
and if we take $f$ to be a constant, we find
\begin{widetext}
\be 
\xi'=\pm v_x C_1 r^2 \sqrt{\frac{\eta'^2+r'^2}{(1-r^2 v_x^2f)[C_1^2 r^4- R^4 (1-r^2 v_x^2f)]}}.
\ee
\end{widetext}
The equation for $r$ is
\be 
-\d_\s \left(\frac{R^4 v_x^2 r'}{r^4 {\cal L}}\right)=\frac{\d \cal L}{\d r}\; .\label{r-eqt}
\ee
\ese

In order to simplify this system, one can try to set one further constraint, $\eta=constant$. From the equation of motion for $\xi$, we find
\bse
\be 
\xi'=\pm \frac{r^2 C_1 v_x r'}{\sqrt{(1-fr^2v_x^2)[r^4 C_1^2 - R^4 (1-fr^2v_x^2)]}}.
\ee
\ese
We see that if we set $r=\s$, we find an inconsistent configuration, since from this last equation
\bse
\be 
\xi'=\pm \frac{r^2 C_1 v_x}{\sqrt{(1-fr^2v_x^2)[r^4 C_1^2 - R^4 (1-fr^2v_x^2)]}}\; ,
\ee
while from the equation of motion for $r$ we have
\be 
\left(\frac{R^4(1-r^2 v_x^2f)}{r^4}\xi'^2-C_3^2 \right)\xi'^2=\frac{-C_3^2 v^2_x }{(1-r^2 v_x^2f)}\; .
\ee
\ese
By substitution we can see that both equations are not consistent, unless $C_1=C_3=0$, where the constant
\be 
C_3=\frac{R^4v_x^2}{r^4 {\cal L}}+{\cal L}\; 
\ee
solves the equation (\ref{r-eqt}) for $r=\s$ and $\eta'= 0$.

On the other hand, we can take the system with $\eta'\neq 0$, but with $r=\s$. The equations of motion are
\bse 
\be 
\eta'=\pm \frac{C_0 r^2}{v_x}\sqrt{\frac{(1-r^2 v_x^2 f)\xi'^2-v_x^2}{R^4 v_x^2+C_0^2r^4}},\label{constraint}
\ee
and
\begin{widetext}
\be 
\xi'=\pm v_x C_1 r^2 \sqrt{\frac{\eta'^2+1}{(1-r^2 v_x^2f)[C_1^2  r^4- R^4 (1-r^2 v_x^2f)]}}.
\ee
\end{widetext}
\ese
Finally, we see that the integration constant $C_0$ is equal to the conserved charge $\Pi_\eta=\frac{\d \cal L}{\d\eta'}$, therefore, the drag force of this configuration is
\be 
F_{drag}=-\frac{R^4 v_x^2 \eta'}{r^4\cal L}\; ,
\ee
but in order for this to be well defined, we need to solve the equation of motion for the coordinate $\xi$ -- probably numerically -- and we also need to consider the additional condition $(1-v_x^2f)\xi'^2-v_x^2>0$, that comes from the reality condition of the equation (\ref{constraint}). Therefore, we can see that in this case we can find nontrivial drag forces at zero temperature.

\section{Conclusions}

In this paper we have reconsidered some string configurations that give Wilson loops in the dual nonrelativistic field theory. In summary, we studied strings moving in spacetimes with Schr\"odinger and Lifshitz symmetries.

We started with Schr\"odinger spacetimes, and reviewed the string configurations with constant compact dimensions \cite{Siahaan:2011sw}. In this case, we have some physical configurations and we calculated the quark-antiquark distance and potential. By extension, we considered the string moving along the $x$-direction and we calculated a nonzero drag force for such a configuration.

Taking into account the motion along the compact extra dimension, $\xi=\xi(\tau)$, we reconsidered the configuration of \cite{Kluson:2009vy}. We pointed that one cannot claim that this configuration is unphysical yet; in fact, there are some issues that must be taken into account: first, the role of the compact coordinate $\xi$ is not clear, and we need to remember that there are genuine singularities at the end of the space. Also, at the present stage of development, we can consider a parity transformation $\xi\to -\xi$, and, apparently, this transformation makes the system well defined. But it is obviously a problem to be scrutinized.

Alternatively, we pointed out that the coordinate $\xi$ is compact, then the configuration with dependence $\xi=\xi(\s)$ may make physical sense. In this case, we found a coupled system of differential equations.

For the Lifshitz case, we saw that there are some cases in which the analysis is the same as in the Schr\"odinger solution for constant compact dimension \cite{Danielsson:2009gi}. On the other hand, we have considered the Lifshitz solution related to the construction given by \cite{Donos:2010tu}, and we saw that a rich scenario emerges. For the case with constant compact direction the solution is unphysical. 

For a compact dimension $\xi$ with dependence on the dimension $\tau$, we have calculated the quark-antiquark potential in \cite{Araujo:2015dba}. Finally, for the compact dimension with dependence on the coordinate $\s$, we calculated the drag force of the string moving through this background.

We recall that we must be careful in using these nonrelativistic spaces. An interesting question is whether the systems of differential equations have solutions or not. It is also very promising to consider the effect of fields of the NS-NS sector on the string, or quantum effects similar to \cite{Sonnenschein:1999if}. We hope to return to some of these points in a future work.

\begin{acknowledgments}
The work of T.A. is supported by CNPq Grant No. 140588/2012-4. The author would like to thank Prieslei Goulart, for reading the final version of this work. Also, the author is grateful to Horatiu Nastase and Yang Lei for stimulating discussions.
\end{acknowledgments}



\bibliography{myref}{}

\begin{thebibliography}{45}%
\makeatletter
\providecommand \@ifxundefined [1]{%
 \@ifx{#1\undefined}
}%
\providecommand \@ifnum [1]{%
 \ifnum #1\expandafter \@firstoftwo
 \else \expandafter \@secondoftwo
 \fi
}%
\providecommand \@ifx [1]{%
 \ifx #1\expandafter \@firstoftwo
 \else \expandafter \@secondoftwo
 \fi
}%
\providecommand \natexlab [1]{#1}%
\providecommand \enquote  [1]{``#1''}%
\providecommand \bibnamefont  [1]{#1}%
\providecommand \bibfnamefont [1]{#1}%
\providecommand \citenamefont [1]{#1}%
\providecommand \href@noop [0]{\@secondoftwo}%
\providecommand \href [0]{\begingroup \@sanitize@url \@href}%
\providecommand \@href[1]{\@@startlink{#1}\@@href}%
\providecommand \@@href[1]{\endgroup#1\@@endlink}%
\providecommand \@sanitize@url [0]{\catcode `\\12\catcode `\$12\catcode
  `\&12\catcode `\#12\catcode `\^12\catcode `\_12\catcode `\%12\relax}%
\providecommand \@@startlink[1]{}%
\providecommand \@@endlink[0]{}%
\providecommand \url  [0]{\begingroup\@sanitize@url \@url }%
\providecommand \@url [1]{\endgroup\@href {#1}{\urlprefix }}%
\providecommand \urlprefix  [0]{URL }%
\providecommand \Eprint [0]{\href }%
\providecommand \doibase [0]{http://dx.doi.org/}%
\providecommand \selectlanguage [0]{\@gobble}%
\providecommand \bibinfo  [0]{\@secondoftwo}%
\providecommand \bibfield  [0]{\@secondoftwo}%
\providecommand \translation [1]{[#1]}%
\providecommand \BibitemOpen [0]{}%
\providecommand \bibitemStop [0]{}%
\providecommand \bibitemNoStop [0]{.\EOS\space}%
\providecommand \EOS [0]{\spacefactor3000\relax}%
\providecommand \BibitemShut  [1]{\csname bibitem#1\endcsname}%
\let\auto@bib@innerbib\@empty
\bibitem [{\citenamefont {Maldacena}(1999)}]{Maldacena:1997re}%
  \BibitemOpen
  \bibfield  {author} {\bibinfo {author} {\bibfnamefont {J.~M.}\ \bibnamefont
  {Maldacena}},\ }\href {\doibase 10.1023/A:1026654312961} {\bibfield
  {journal} {\bibinfo  {journal} {Int.J.Theor.Phys.}\ }\textbf {\bibinfo
  {volume} {38}},\ \bibinfo {pages} {1113} (\bibinfo {year} {1999})},\ \Eprint
  {http://arxiv.org/abs/hep-th/9711200} {arXiv:hep-th/9711200 [hep-th]}
  \BibitemShut {NoStop}%
\bibitem [{\citenamefont {Witten}(1998)}]{Witten:1998qj}%
  \BibitemOpen
  \bibfield  {author} {\bibinfo {author} {\bibfnamefont {E.}~\bibnamefont
  {Witten}},\ }\href@noop {} {\bibfield  {journal} {\bibinfo  {journal}
  {Adv.Theor.Math.Phys.}\ }\textbf {\bibinfo {volume} {2}},\ \bibinfo {pages}
  {253} (\bibinfo {year} {1998})},\ \Eprint
  {http://arxiv.org/abs/hep-th/9802150} {arXiv:hep-th/9802150 [hep-th]}
  \BibitemShut {NoStop}%
\bibitem [{\citenamefont {Smilga}(2001)}]{Smilga2001}%
  \BibitemOpen
  \bibfield  {author} {\bibinfo {author} {\bibfnamefont {A.~V.}\ \bibnamefont
  {Smilga}},\ }\href@noop {} {\emph {\bibinfo {title} {Lectures on Quantum
  Chromodynamics}}}\ (\bibinfo  {publisher} {World Scientific,},\ \bibinfo
  {year} {2001})\BibitemShut {NoStop}%
\bibitem [{\citenamefont {Makeenko}(2010)}]{Makeenko:2009dw}%
  \BibitemOpen
  \bibfield  {author} {\bibinfo {author} {\bibfnamefont {Y.}~\bibnamefont
  {Makeenko}},\ }\bibfield  {booktitle} {\emph {\bibinfo {booktitle} {{12th
  International Moscow School of Physics and 37th ITEP Winter School of Physics
  Moscow, Russia, February 9-16, 2009}}},\ }\href {\doibase
  10.1134/S106377881005011X} {\bibfield  {journal} {\bibinfo  {journal} {Phys.
  Atom. Nucl.}\ }\textbf {\bibinfo {volume} {73}},\ \bibinfo {pages} {878}
  (\bibinfo {year} {2010})},\ \Eprint {http://arxiv.org/abs/0906.4487}
  {arXiv:0906.4487 [hep-th]} \BibitemShut {NoStop}%
\bibitem [{\citenamefont {Maldacena}(1998)}]{Maldacena:1998im}%
  \BibitemOpen
  \bibfield  {author} {\bibinfo {author} {\bibfnamefont {J.~M.}\ \bibnamefont
  {Maldacena}},\ }\href {\doibase 10.1103/PhysRevLett.80.4859} {\bibfield
  {journal} {\bibinfo  {journal} {Phys. Rev. Lett.}\ }\textbf {\bibinfo
  {volume} {80}},\ \bibinfo {pages} {4859} (\bibinfo {year} {1998})},\ \Eprint
  {http://arxiv.org/abs/hep-th/9803002} {arXiv:hep-th/9803002 [hep-th]}
  \BibitemShut {NoStop}%
\bibitem [{\citenamefont {Drukker}\ \emph {et~al.}(1999)\citenamefont
  {Drukker}, \citenamefont {Gross},\ and\ \citenamefont
  {Ooguri}}]{Drukker:1999zq}%
  \BibitemOpen
  \bibfield  {author} {\bibinfo {author} {\bibfnamefont {N.}~\bibnamefont
  {Drukker}}, \bibinfo {author} {\bibfnamefont {D.~J.}\ \bibnamefont {Gross}},
  \ and\ \bibinfo {author} {\bibfnamefont {H.}~\bibnamefont {Ooguri}},\ }\href
  {\doibase 10.1103/PhysRevD.60.125006} {\bibfield  {journal} {\bibinfo
  {journal} {Phys. Rev.}\ }\textbf {\bibinfo {volume} {D60}},\ \bibinfo {pages}
  {125006} (\bibinfo {year} {1999})},\ \Eprint
  {http://arxiv.org/abs/hep-th/9904191} {arXiv:hep-th/9904191 [hep-th]}
  \BibitemShut {NoStop}%
\bibitem [{\citenamefont {Sonnenschein}(1999)}]{Sonnenschein:1999if}%
  \BibitemOpen
  \bibfield  {author} {\bibinfo {author} {\bibfnamefont {J.}~\bibnamefont
  {Sonnenschein}},\ }\href@noop {} {\ ,\ \bibinfo {pages} {219} (\bibinfo
  {year} {1999})},\ \Eprint {http://arxiv.org/abs/hep-th/0003032}
  {arXiv:hep-th/0003032 [hep-th]} \BibitemShut {NoStop}%
\bibitem [{\citenamefont {Nunez}\ \emph {et~al.}(2010)\citenamefont {Nunez},
  \citenamefont {Piai},\ and\ \citenamefont {Rago}}]{Nunez:2009da}%
  \BibitemOpen
  \bibfield  {author} {\bibinfo {author} {\bibfnamefont {C.}~\bibnamefont
  {Nunez}}, \bibinfo {author} {\bibfnamefont {M.}~\bibnamefont {Piai}}, \ and\
  \bibinfo {author} {\bibfnamefont {A.}~\bibnamefont {Rago}},\ }\href {\doibase
  10.1103/PhysRevD.81.086001} {\bibfield  {journal} {\bibinfo  {journal}
  {Phys.Rev.}\ }\textbf {\bibinfo {volume} {D81}},\ \bibinfo {pages} {086001}
  (\bibinfo {year} {2010})},\ \Eprint {http://arxiv.org/abs/0909.0748}
  {arXiv:0909.0748 [hep-th]} \BibitemShut {NoStop}%
\bibitem [{\citenamefont {Aharony}\ \emph {et~al.}(2000)\citenamefont
  {Aharony}, \citenamefont {Gubser}, \citenamefont {Maldacena}, \citenamefont
  {Ooguri},\ and\ \citenamefont {Oz}}]{Aharony:1999ti}%
  \BibitemOpen
  \bibfield  {author} {\bibinfo {author} {\bibfnamefont {O.}~\bibnamefont
  {Aharony}}, \bibinfo {author} {\bibfnamefont {S.~S.}\ \bibnamefont {Gubser}},
  \bibinfo {author} {\bibfnamefont {J.~M.}\ \bibnamefont {Maldacena}}, \bibinfo
  {author} {\bibfnamefont {H.}~\bibnamefont {Ooguri}}, \ and\ \bibinfo {author}
  {\bibfnamefont {Y.}~\bibnamefont {Oz}},\ }\href {\doibase
  10.1016/S0370-1573(99)00083-6} {\bibfield  {journal} {\bibinfo  {journal}
  {Phys.Rept.}\ }\textbf {\bibinfo {volume} {323}},\ \bibinfo {pages} {183}
  (\bibinfo {year} {2000})},\ \Eprint {http://arxiv.org/abs/hep-th/9905111}
  {arXiv:hep-th/9905111 [hep-th]} \BibitemShut {NoStop}%
\bibitem [{\citenamefont {Polchinski}(2010)}]{Polchinski:2010hw}%
  \BibitemOpen
  \bibfield  {author} {\bibinfo {author} {\bibfnamefont {J.}~\bibnamefont
  {Polchinski}},\ }\href {\doibase 10.1142/9789814350525_0001} {\ ,\ \bibinfo
  {pages} {3} (\bibinfo {year} {2010})},\ \Eprint
  {http://arxiv.org/abs/1010.6134} {arXiv:1010.6134 [hep-th]} \BibitemShut
  {NoStop}%
\bibitem [{\citenamefont {Gomis}\ and\ \citenamefont
  {Ooguri}(2001)}]{Gomis:2000bd}%
  \BibitemOpen
  \bibfield  {author} {\bibinfo {author} {\bibfnamefont {J.}~\bibnamefont
  {Gomis}}\ and\ \bibinfo {author} {\bibfnamefont {H.}~\bibnamefont {Ooguri}},\
  }\href {\doibase 10.1063/1.1372697} {\bibfield  {journal} {\bibinfo
  {journal} {J. Math. Phys.}\ }\textbf {\bibinfo {volume} {42}},\ \bibinfo
  {pages} {3127} (\bibinfo {year} {2001})},\ \Eprint
  {http://arxiv.org/abs/hep-th/0009181} {arXiv:hep-th/0009181 [hep-th]}
  \BibitemShut {NoStop}%
\bibitem [{\citenamefont {Nishida}\ and\ \citenamefont
  {Son}(2007)}]{Nishida:2007pj}%
  \BibitemOpen
  \bibfield  {author} {\bibinfo {author} {\bibfnamefont {Y.}~\bibnamefont
  {Nishida}}\ and\ \bibinfo {author} {\bibfnamefont {D.~T.}\ \bibnamefont
  {Son}},\ }\href {\doibase 10.1103/PhysRevD.76.086004} {\bibfield  {journal}
  {\bibinfo  {journal} {Phys.Rev.}\ }\textbf {\bibinfo {volume} {D76}},\
  \bibinfo {pages} {086004} (\bibinfo {year} {2007})},\ \Eprint
  {http://arxiv.org/abs/0706.3746} {arXiv:0706.3746 [hep-th]} \BibitemShut
  {NoStop}%
\bibitem [{\citenamefont {Son}(2008)}]{Son:2008ye}%
  \BibitemOpen
  \bibfield  {author} {\bibinfo {author} {\bibfnamefont {D.}~\bibnamefont
  {Son}},\ }\href {\doibase 10.1103/PhysRevD.78.046003} {\bibfield  {journal}
  {\bibinfo  {journal} {Phys.Rev.}\ }\textbf {\bibinfo {volume} {D78}},\
  \bibinfo {pages} {046003} (\bibinfo {year} {2008})},\ \Eprint
  {http://arxiv.org/abs/0804.3972} {arXiv:0804.3972 [hep-th]} \BibitemShut
  {NoStop}%
\bibitem [{\citenamefont {Balasubramanian}\ and\ \citenamefont
  {McGreevy}(2008)}]{Balasubramanian:2008dm}%
  \BibitemOpen
  \bibfield  {author} {\bibinfo {author} {\bibfnamefont {K.}~\bibnamefont
  {Balasubramanian}}\ and\ \bibinfo {author} {\bibfnamefont {J.}~\bibnamefont
  {McGreevy}},\ }\href {\doibase 10.1103/PhysRevLett.101.061601} {\bibfield
  {journal} {\bibinfo  {journal} {Phys.Rev.Lett.}\ }\textbf {\bibinfo {volume}
  {101}},\ \bibinfo {pages} {061601} (\bibinfo {year} {2008})},\ \Eprint
  {http://arxiv.org/abs/0804.4053} {arXiv:0804.4053 [hep-th]} \BibitemShut
  {NoStop}%
\bibitem [{\citenamefont {Herzog}\ \emph {et~al.}(2008)\citenamefont {Herzog},
  \citenamefont {Rangamani},\ and\ \citenamefont {Ross}}]{Herzog:2008wg}%
  \BibitemOpen
  \bibfield  {author} {\bibinfo {author} {\bibfnamefont {C.~P.}\ \bibnamefont
  {Herzog}}, \bibinfo {author} {\bibfnamefont {M.}~\bibnamefont {Rangamani}}, \
  and\ \bibinfo {author} {\bibfnamefont {S.~F.}\ \bibnamefont {Ross}},\ }\href
  {\doibase 10.1088/1126-6708/2008/11/080} {\bibfield  {journal} {\bibinfo
  {journal} {JHEP}\ }\textbf {\bibinfo {volume} {0811}},\ \bibinfo {pages}
  {080} (\bibinfo {year} {2008})},\ \Eprint {http://arxiv.org/abs/0807.1099}
  {arXiv:0807.1099 [hep-th]} \BibitemShut {NoStop}%
\bibitem [{\citenamefont {Maldacena}\ \emph {et~al.}(2008)\citenamefont
  {Maldacena}, \citenamefont {Martelli},\ and\ \citenamefont
  {Tachikawa}}]{Maldacena:2008wh}%
  \BibitemOpen
  \bibfield  {author} {\bibinfo {author} {\bibfnamefont {J.}~\bibnamefont
  {Maldacena}}, \bibinfo {author} {\bibfnamefont {D.}~\bibnamefont {Martelli}},
  \ and\ \bibinfo {author} {\bibfnamefont {Y.}~\bibnamefont {Tachikawa}},\
  }\href {\doibase 10.1088/1126-6708/2008/10/072} {\bibfield  {journal}
  {\bibinfo  {journal} {JHEP}\ }\textbf {\bibinfo {volume} {0810}},\ \bibinfo
  {pages} {072} (\bibinfo {year} {2008})},\ \Eprint
  {http://arxiv.org/abs/0807.1100} {arXiv:0807.1100 [hep-th]} \BibitemShut
  {NoStop}%
\bibitem [{\citenamefont {Adams}\ \emph {et~al.}(2008)\citenamefont {Adams},
  \citenamefont {Balasubramanian},\ and\ \citenamefont
  {McGreevy}}]{Adams:2008wt}%
  \BibitemOpen
  \bibfield  {author} {\bibinfo {author} {\bibfnamefont {A.}~\bibnamefont
  {Adams}}, \bibinfo {author} {\bibfnamefont {K.}~\bibnamefont
  {Balasubramanian}}, \ and\ \bibinfo {author} {\bibfnamefont {J.}~\bibnamefont
  {McGreevy}},\ }\href {\doibase 10.1088/1126-6708/2008/11/059} {\bibfield
  {journal} {\bibinfo  {journal} {JHEP}\ }\textbf {\bibinfo {volume} {0811}},\
  \bibinfo {pages} {059} (\bibinfo {year} {2008})},\ \Eprint
  {http://arxiv.org/abs/0807.1111} {arXiv:0807.1111 [hep-th]} \BibitemShut
  {NoStop}%
\bibitem [{\citenamefont {Ko}\ \emph {et~al.}(2015)\citenamefont {Ko},
  \citenamefont {Melby-Thompson}, \citenamefont {Meyer},\ and\ \citenamefont
  {Park}}]{Ko:2015rha}%
  \BibitemOpen
  \bibfield  {author} {\bibinfo {author} {\bibfnamefont {S.~M.}\ \bibnamefont
  {Ko}}, \bibinfo {author} {\bibfnamefont {C.}~\bibnamefont {Melby-Thompson}},
  \bibinfo {author} {\bibfnamefont {R.}~\bibnamefont {Meyer}}, \ and\ \bibinfo
  {author} {\bibfnamefont {J.-H.}\ \bibnamefont {Park}},\ }\href@noop {} {\
  (\bibinfo {year} {2015})},\ \Eprint {http://arxiv.org/abs/1508.01121}
  {arXiv:1508.01121 [hep-th]} \BibitemShut {NoStop}%
\bibitem [{\citenamefont {Hartnoll}(2009)}]{Hartnoll:2009sz}%
  \BibitemOpen
  \bibfield  {author} {\bibinfo {author} {\bibfnamefont {S.~A.}\ \bibnamefont
  {Hartnoll}},\ }\href {\doibase 10.1088/0264-9381/26/22/224002} {\bibfield
  {journal} {\bibinfo  {journal} {Class.Quant.Grav.}\ }\textbf {\bibinfo
  {volume} {26}},\ \bibinfo {pages} {224002} (\bibinfo {year} {2009})},\
  \Eprint {http://arxiv.org/abs/0903.3246} {arXiv:0903.3246 [hep-th]}
  \BibitemShut {NoStop}%
\bibitem [{\citenamefont {Donos}\ and\ \citenamefont
  {Gauntlett}(2009)}]{Donos:2009xc}%
  \BibitemOpen
  \bibfield  {author} {\bibinfo {author} {\bibfnamefont {A.}~\bibnamefont
  {Donos}}\ and\ \bibinfo {author} {\bibfnamefont {J.~P.}\ \bibnamefont
  {Gauntlett}},\ }\href {\doibase 10.1088/1126-6708/2009/07/042} {\bibfield
  {journal} {\bibinfo  {journal} {JHEP}\ }\textbf {\bibinfo {volume} {0907}},\
  \bibinfo {pages} {042} (\bibinfo {year} {2009})},\ \Eprint
  {http://arxiv.org/abs/0905.1098} {arXiv:0905.1098 [hep-th]} \BibitemShut
  {NoStop}%
\bibitem [{\citenamefont {Donos}\ and\ \citenamefont
  {Gauntlett}(2010)}]{Donos:2010tu}%
  \BibitemOpen
  \bibfield  {author} {\bibinfo {author} {\bibfnamefont {A.}~\bibnamefont
  {Donos}}\ and\ \bibinfo {author} {\bibfnamefont {J.~P.}\ \bibnamefont
  {Gauntlett}},\ }\href {\doibase 10.1007/JHEP12(2010)002} {\bibfield
  {journal} {\bibinfo  {journal} {JHEP}\ }\textbf {\bibinfo {volume} {1012}},\
  \bibinfo {pages} {002} (\bibinfo {year} {2010})},\ \Eprint
  {http://arxiv.org/abs/1008.2062} {arXiv:1008.2062 [hep-th]} \BibitemShut
  {NoStop}%
\bibitem [{\citenamefont {Balasubramanian}\ and\ \citenamefont
  {Narayan}(2010)}]{Balasubramanian:2010uk}%
  \BibitemOpen
  \bibfield  {author} {\bibinfo {author} {\bibfnamefont {K.}~\bibnamefont
  {Balasubramanian}}\ and\ \bibinfo {author} {\bibfnamefont {K.}~\bibnamefont
  {Narayan}},\ }\href {\doibase 10.1007/JHEP08(2010)014} {\bibfield  {journal}
  {\bibinfo  {journal} {JHEP}\ }\textbf {\bibinfo {volume} {1008}},\ \bibinfo
  {pages} {014} (\bibinfo {year} {2010})},\ \Eprint
  {http://arxiv.org/abs/1005.3291} {arXiv:1005.3291 [hep-th]} \BibitemShut
  {NoStop}%
\bibitem [{\citenamefont {Gregory}\ \emph {et~al.}(2010)\citenamefont
  {Gregory}, \citenamefont {Parameswaran}, \citenamefont {Tasinato},\ and\
  \citenamefont {Zavala}}]{Gregory:2010gx}%
  \BibitemOpen
  \bibfield  {author} {\bibinfo {author} {\bibfnamefont {R.}~\bibnamefont
  {Gregory}}, \bibinfo {author} {\bibfnamefont {S.~L.}\ \bibnamefont
  {Parameswaran}}, \bibinfo {author} {\bibfnamefont {G.}~\bibnamefont
  {Tasinato}}, \ and\ \bibinfo {author} {\bibfnamefont {I.}~\bibnamefont
  {Zavala}},\ }\href {\doibase 10.1007/JHEP12(2010)047} {\bibfield  {journal}
  {\bibinfo  {journal} {JHEP}\ }\textbf {\bibinfo {volume} {12}},\ \bibinfo
  {pages} {047} (\bibinfo {year} {2010})},\ \Eprint
  {http://arxiv.org/abs/1009.3445} {arXiv:1009.3445 [hep-th]} \BibitemShut
  {NoStop}%
\bibitem [{\citenamefont {Gubser}(2006)}]{Gubser:2006bz}%
  \BibitemOpen
  \bibfield  {author} {\bibinfo {author} {\bibfnamefont {S.~S.}\ \bibnamefont
  {Gubser}},\ }\href {\doibase 10.1103/PhysRevD.74.126005} {\bibfield
  {journal} {\bibinfo  {journal} {Phys.Rev.}\ }\textbf {\bibinfo {volume}
  {D74}},\ \bibinfo {pages} {126005} (\bibinfo {year} {2006})},\ \Eprint
  {http://arxiv.org/abs/hep-th/0605182} {arXiv:hep-th/0605182 [hep-th]}
  \BibitemShut {NoStop}%
\bibitem [{\citenamefont {Chernicoff}\ \emph {et~al.}(2006)\citenamefont
  {Chernicoff}, \citenamefont {Garcia},\ and\ \citenamefont
  {Guijosa}}]{Chernicoff:2006hi}%
  \BibitemOpen
  \bibfield  {author} {\bibinfo {author} {\bibfnamefont {M.}~\bibnamefont
  {Chernicoff}}, \bibinfo {author} {\bibfnamefont {J.~A.}\ \bibnamefont
  {Garcia}}, \ and\ \bibinfo {author} {\bibfnamefont {A.}~\bibnamefont
  {Guijosa}},\ }\href {\doibase 10.1088/1126-6708/2006/09/068} {\bibfield
  {journal} {\bibinfo  {journal} {JHEP}\ }\textbf {\bibinfo {volume} {09}},\
  \bibinfo {pages} {068} (\bibinfo {year} {2006})},\ \Eprint
  {http://arxiv.org/abs/hep-th/0607089} {arXiv:hep-th/0607089 [hep-th]}
  \BibitemShut {NoStop}%
\bibitem [{\citenamefont {Herzog}(2006)}]{Herzog:2006se}%
  \BibitemOpen
  \bibfield  {author} {\bibinfo {author} {\bibfnamefont {C.~P.}\ \bibnamefont
  {Herzog}},\ }\href {\doibase 10.1088/1126-6708/2006/09/032} {\bibfield
  {journal} {\bibinfo  {journal} {JHEP}\ }\textbf {\bibinfo {volume} {0609}},\
  \bibinfo {pages} {032} (\bibinfo {year} {2006})},\ \Eprint
  {http://arxiv.org/abs/hep-th/0605191} {arXiv:hep-th/0605191 [hep-th]}
  \BibitemShut {NoStop}%
\bibitem [{\citenamefont {Herzog}\ \emph {et~al.}(2006)\citenamefont {Herzog},
  \citenamefont {Karch}, \citenamefont {Kovtun}, \citenamefont {Kozcaz},\ and\
  \citenamefont {Yaffe}}]{Herzog:2006gh}%
  \BibitemOpen
  \bibfield  {author} {\bibinfo {author} {\bibfnamefont {C.~P.}\ \bibnamefont
  {Herzog}}, \bibinfo {author} {\bibfnamefont {A.}~\bibnamefont {Karch}},
  \bibinfo {author} {\bibfnamefont {P.}~\bibnamefont {Kovtun}}, \bibinfo
  {author} {\bibfnamefont {C.}~\bibnamefont {Kozcaz}}, \ and\ \bibinfo {author}
  {\bibfnamefont {L.~G.}\ \bibnamefont {Yaffe}},\ }\href {\doibase
  10.1088/1126-6708/2006/07/013} {\bibfield  {journal} {\bibinfo  {journal}
  {JHEP}\ }\textbf {\bibinfo {volume} {07}},\ \bibinfo {pages} {013} (\bibinfo
  {year} {2006})},\ \Eprint {http://arxiv.org/abs/hep-th/0605158}
  {arXiv:hep-th/0605158 [hep-th]} \BibitemShut {NoStop}%
\bibitem [{\citenamefont {Horowitz}\ and\ \citenamefont
  {Ross}(1997)}]{Horowitz:1997uc}%
  \BibitemOpen
  \bibfield  {author} {\bibinfo {author} {\bibfnamefont {G.~T.}\ \bibnamefont
  {Horowitz}}\ and\ \bibinfo {author} {\bibfnamefont {S.~F.}\ \bibnamefont
  {Ross}},\ }\href {\doibase 10.1103/PhysRevD.56.2180} {\bibfield  {journal}
  {\bibinfo  {journal} {Phys. Rev.}\ }\textbf {\bibinfo {volume} {D56}},\
  \bibinfo {pages} {2180} (\bibinfo {year} {1997})},\ \Eprint
  {http://arxiv.org/abs/hep-th/9704058} {arXiv:hep-th/9704058 [hep-th]}
  \BibitemShut {NoStop}%
\bibitem [{\citenamefont {Akhavan}\ \emph {et~al.}(2009)\citenamefont
  {Akhavan}, \citenamefont {Alishahiha}, \citenamefont {Davody},\ and\
  \citenamefont {Vahedi}}]{Akhavan:2008ep}%
  \BibitemOpen
  \bibfield  {author} {\bibinfo {author} {\bibfnamefont {A.}~\bibnamefont
  {Akhavan}}, \bibinfo {author} {\bibfnamefont {M.}~\bibnamefont {Alishahiha}},
  \bibinfo {author} {\bibfnamefont {A.}~\bibnamefont {Davody}}, \ and\ \bibinfo
  {author} {\bibfnamefont {A.}~\bibnamefont {Vahedi}},\ }\href {\doibase
  10.1088/1126-6708/2009/03/053} {\bibfield  {journal} {\bibinfo  {journal}
  {JHEP}\ }\textbf {\bibinfo {volume} {0903}},\ \bibinfo {pages} {053}
  (\bibinfo {year} {2009})},\ \Eprint {http://arxiv.org/abs/0811.3067}
  {arXiv:0811.3067 [hep-th]} \BibitemShut {NoStop}%
\bibitem [{\citenamefont {Kluson}(2010)}]{Kluson:2009vy}%
  \BibitemOpen
  \bibfield  {author} {\bibinfo {author} {\bibfnamefont {J.}~\bibnamefont
  {Kluson}},\ }\href {\doibase 10.1103/PhysRevD.81.106006} {\bibfield
  {journal} {\bibinfo  {journal} {Phys.Rev.}\ }\textbf {\bibinfo {volume}
  {D81}},\ \bibinfo {pages} {106006} (\bibinfo {year} {2010})},\ \Eprint
  {http://arxiv.org/abs/0912.4587} {arXiv:0912.4587 [hep-th]} \BibitemShut
  {NoStop}%
\bibitem [{\citenamefont {Siahaan}(2011)}]{Siahaan:2011sw}%
  \BibitemOpen
  \bibfield  {author} {\bibinfo {author} {\bibfnamefont {H.~M.}\ \bibnamefont
  {Siahaan}},\ }\href {\doibase 10.1142/S0217732311036255} {\bibfield
  {journal} {\bibinfo  {journal} {Mod.Phys.Lett.}\ }\textbf {\bibinfo {volume}
  {A26}},\ \bibinfo {pages} {1719} (\bibinfo {year} {2011})},\ \Eprint
  {http://arxiv.org/abs/1106.5008} {arXiv:1106.5008 [hep-th]} \BibitemShut
  {NoStop}%
\bibitem [{\citenamefont {Kinar}\ \emph {et~al.}(2000)\citenamefont {Kinar},
  \citenamefont {Schreiber},\ and\ \citenamefont
  {Sonnenschein}}]{Kinar:1998vq}%
  \BibitemOpen
  \bibfield  {author} {\bibinfo {author} {\bibfnamefont {Y.}~\bibnamefont
  {Kinar}}, \bibinfo {author} {\bibfnamefont {E.}~\bibnamefont {Schreiber}}, \
  and\ \bibinfo {author} {\bibfnamefont {J.}~\bibnamefont {Sonnenschein}},\
  }\href {\doibase 10.1016/S0550-3213(99)00652-5} {\bibfield  {journal}
  {\bibinfo  {journal} {Nucl. Phys.}\ }\textbf {\bibinfo {volume} {B566}},\
  \bibinfo {pages} {103} (\bibinfo {year} {2000})},\ \Eprint
  {http://arxiv.org/abs/hep-th/9811192} {arXiv:hep-th/9811192 [hep-th]}
  \BibitemShut {NoStop}%
\bibitem [{\citenamefont {Araujo}\ and\ \citenamefont
  {Nastase}(2015)}]{Araujo:2015dba}%
  \BibitemOpen
  \bibfield  {author} {\bibinfo {author} {\bibfnamefont {T.~R.}\ \bibnamefont
  {Araujo}}\ and\ \bibinfo {author} {\bibfnamefont {H.}~\bibnamefont
  {Nastase}},\ }\href@noop {} {\  (\bibinfo {year} {2015})},\ \Eprint
  {http://arxiv.org/abs/1508.06568} {arXiv:1508.06568 [hep-th]} \BibitemShut
  {NoStop}%
\bibitem [{Note1()}]{Note1}%
  \BibitemOpen
  \bibinfo {note} {Using the definition of the Gamma function to extend the
  domain of the Beta function.}\BibitemShut {Stop}%
\bibitem [{\citenamefont {Bachas}(1986)}]{Bachas:1985xs}%
  \BibitemOpen
  \bibfield  {author} {\bibinfo {author} {\bibfnamefont {C.}~\bibnamefont
  {Bachas}},\ }\href {\doibase 10.1103/PhysRevD.33.2723} {\bibfield  {journal}
  {\bibinfo  {journal} {Phys.Rev.}\ }\textbf {\bibinfo {volume} {D33}},\
  \bibinfo {pages} {2723} (\bibinfo {year} {1986})}\BibitemShut {NoStop}%
\bibitem [{\citenamefont {Arias}\ and\ \citenamefont
  {Silva}(2010)}]{Arias:2009me}%
  \BibitemOpen
  \bibfield  {author} {\bibinfo {author} {\bibfnamefont {R.~E.}\ \bibnamefont
  {Arias}}\ and\ \bibinfo {author} {\bibfnamefont {G.~A.}\ \bibnamefont
  {Silva}},\ }\href {\doibase 10.1007/JHEP01(2010)023} {\bibfield  {journal}
  {\bibinfo  {journal} {JHEP}\ }\textbf {\bibinfo {volume} {1001}},\ \bibinfo
  {pages} {023} (\bibinfo {year} {2010})},\ \Eprint
  {http://arxiv.org/abs/0911.0662} {arXiv:0911.0662 [hep-th]} \BibitemShut
  {NoStop}%
\bibitem [{\citenamefont {Gubser}\ \emph {et~al.}(1998)\citenamefont {Gubser},
  \citenamefont {Klebanov},\ and\ \citenamefont {Polyakov}}]{Gubser:1998bc}%
  \BibitemOpen
  \bibfield  {author} {\bibinfo {author} {\bibfnamefont {S.}~\bibnamefont
  {Gubser}}, \bibinfo {author} {\bibfnamefont {I.~R.}\ \bibnamefont
  {Klebanov}}, \ and\ \bibinfo {author} {\bibfnamefont {A.~M.}\ \bibnamefont
  {Polyakov}},\ }\href {\doibase 10.1016/S0370-2693(98)00377-3} {\bibfield
  {journal} {\bibinfo  {journal} {Phys.Lett.}\ }\textbf {\bibinfo {volume}
  {B428}},\ \bibinfo {pages} {105} (\bibinfo {year} {1998})},\ \Eprint
  {http://arxiv.org/abs/hep-th/9802109} {arXiv:hep-th/9802109 [hep-th]}
  \BibitemShut {NoStop}%
\bibitem [{\citenamefont {Fadafan}(2009)}]{Fadafan:2009an}%
  \BibitemOpen
  \bibfield  {author} {\bibinfo {author} {\bibfnamefont {K.~B.}\ \bibnamefont
  {Fadafan}},\ }\href@noop {} {\  (\bibinfo {year} {2009})},\ \Eprint
  {http://arxiv.org/abs/0912.4873} {arXiv:0912.4873 [hep-th]} \BibitemShut
  {NoStop}%
\bibitem [{\citenamefont {Balasubramanian}\ and\ \citenamefont
  {McGreevy}(2011)}]{Balasubramanian:2010uw}%
  \BibitemOpen
  \bibfield  {author} {\bibinfo {author} {\bibfnamefont {K.}~\bibnamefont
  {Balasubramanian}}\ and\ \bibinfo {author} {\bibfnamefont {J.}~\bibnamefont
  {McGreevy}},\ }\href {\doibase 10.1007/JHEP01(2011)137} {\bibfield  {journal}
  {\bibinfo  {journal} {JHEP}\ }\textbf {\bibinfo {volume} {1101}},\ \bibinfo
  {pages} {137} (\bibinfo {year} {2011})},\ \Eprint
  {http://arxiv.org/abs/1007.2184} {arXiv:1007.2184 [hep-th]} \BibitemShut
  {NoStop}%
\bibitem [{\citenamefont {Blau}\ \emph {et~al.}(2009)\citenamefont {Blau},
  \citenamefont {Hartong},\ and\ \citenamefont {Rollier}}]{Blau:2009gd}%
  \BibitemOpen
  \bibfield  {author} {\bibinfo {author} {\bibfnamefont {M.}~\bibnamefont
  {Blau}}, \bibinfo {author} {\bibfnamefont {J.}~\bibnamefont {Hartong}}, \
  and\ \bibinfo {author} {\bibfnamefont {B.}~\bibnamefont {Rollier}},\ }\href
  {\doibase 10.1088/1126-6708/2009/07/027} {\bibfield  {journal} {\bibinfo
  {journal} {JHEP}\ }\textbf {\bibinfo {volume} {07}},\ \bibinfo {pages} {027}
  (\bibinfo {year} {2009})},\ \Eprint {http://arxiv.org/abs/0904.3304}
  {arXiv:0904.3304 [hep-th]} \BibitemShut {NoStop}%
\bibitem [{\citenamefont {Gubser}(2000)}]{Gubser:2000nd}%
  \BibitemOpen
  \bibfield  {author} {\bibinfo {author} {\bibfnamefont {S.~S.}\ \bibnamefont
  {Gubser}},\ }\href@noop {} {\bibfield  {journal} {\bibinfo  {journal} {Adv.
  Theor. Math. Phys.}\ }\textbf {\bibinfo {volume} {4}},\ \bibinfo {pages}
  {679} (\bibinfo {year} {2000})},\ \Eprint
  {http://arxiv.org/abs/hep-th/0002160} {arXiv:hep-th/0002160 [hep-th]}
  \BibitemShut {NoStop}%
\bibitem [{\citenamefont {Charmousis}\ \emph {et~al.}(2010)\citenamefont
  {Charmousis}, \citenamefont {Gouteraux}, \citenamefont {Kim}, \citenamefont
  {Kiritsis},\ and\ \citenamefont {Meyer}}]{Charmousis:2010zz}%
  \BibitemOpen
  \bibfield  {author} {\bibinfo {author} {\bibfnamefont {C.}~\bibnamefont
  {Charmousis}}, \bibinfo {author} {\bibfnamefont {B.}~\bibnamefont
  {Gouteraux}}, \bibinfo {author} {\bibfnamefont {B.~S.}\ \bibnamefont {Kim}},
  \bibinfo {author} {\bibfnamefont {E.}~\bibnamefont {Kiritsis}}, \ and\
  \bibinfo {author} {\bibfnamefont {R.}~\bibnamefont {Meyer}},\ }\href
  {\doibase 10.1007/JHEP11(2010)151} {\bibfield  {journal} {\bibinfo  {journal}
  {JHEP}\ }\textbf {\bibinfo {volume} {11}},\ \bibinfo {pages} {151} (\bibinfo
  {year} {2010})},\ \Eprint {http://arxiv.org/abs/1005.4690} {arXiv:1005.4690
  [hep-th]} \BibitemShut {NoStop}%
\bibitem [{\citenamefont {Lippert}\ \emph {et~al.}(2015)\citenamefont
  {Lippert}, \citenamefont {Meyer},\ and\ \citenamefont
  {Taliotis}}]{Lippert:2014jma}%
  \BibitemOpen
  \bibfield  {author} {\bibinfo {author} {\bibfnamefont {M.}~\bibnamefont
  {Lippert}}, \bibinfo {author} {\bibfnamefont {R.}~\bibnamefont {Meyer}}, \
  and\ \bibinfo {author} {\bibfnamefont {A.}~\bibnamefont {Taliotis}},\ }\href
  {\doibase 10.1007/JHEP01(2015)023} {\bibfield  {journal} {\bibinfo  {journal}
  {JHEP}\ }\textbf {\bibinfo {volume} {01}},\ \bibinfo {pages} {023} (\bibinfo
  {year} {2015})},\ \Eprint {http://arxiv.org/abs/1409.1369} {arXiv:1409.1369
  [hep-th]} \BibitemShut {NoStop}%
\bibitem [{\citenamefont {Andrade}\ \emph {et~al.}(2014)\citenamefont
  {Andrade}, \citenamefont {Lei},\ and\ \citenamefont
  {Ross}}]{Andrade:2014bsa}%
  \BibitemOpen
  \bibfield  {author} {\bibinfo {author} {\bibfnamefont {T.}~\bibnamefont
  {Andrade}}, \bibinfo {author} {\bibfnamefont {Y.}~\bibnamefont {Lei}}, \ and\
  \bibinfo {author} {\bibfnamefont {S.~F.}\ \bibnamefont {Ross}},\ }\href
  {\doibase 10.1088/0264-9381/31/21/215002} {\bibfield  {journal} {\bibinfo
  {journal} {Class. Quant. Grav.}\ }\textbf {\bibinfo {volume} {31}},\ \bibinfo
  {pages} {215002} (\bibinfo {year} {2014})},\ \Eprint
  {http://arxiv.org/abs/1406.6389} {arXiv:1406.6389 [hep-th]} \BibitemShut
  {NoStop}%
\bibitem [{\citenamefont {Danielsson}\ and\ \citenamefont
  {Thorlacius}(2009)}]{Danielsson:2009gi}%
  \BibitemOpen
  \bibfield  {author} {\bibinfo {author} {\bibfnamefont {U.~H.}\ \bibnamefont
  {Danielsson}}\ and\ \bibinfo {author} {\bibfnamefont {L.}~\bibnamefont
  {Thorlacius}},\ }\href {\doibase 10.1088/1126-6708/2009/03/070} {\bibfield
  {journal} {\bibinfo  {journal} {JHEP}\ }\textbf {\bibinfo {volume} {0903}},\
  \bibinfo {pages} {070} (\bibinfo {year} {2009})},\ \Eprint
  {http://arxiv.org/abs/0812.5088} {arXiv:0812.5088 [hep-th]} \BibitemShut
  {NoStop}%
\end{thebibliography}%

\end{document}